
%
%
\input harvmac

\def\CTPa{\it Center for Theoretical Physics, Department of Physics,
      Texas A\&M University}
\def\CTPb{\it College Station, TX 77843-4242, USA}

\def\HARCa{\it Astroparticle Physics Group,
Houston Advanced Research Center (HARC)}
\def\HARCb{\it The Woodlands, TX 77381, USA}

%
%
%
\def\ie{\hbox{\it i.e.}}     
\def\eg{\hbox{\it e.g.}}

\def\VEV#1{\left\langle #1\right\rangle}
\catcode`\@=11 

\def\lsim{\mathrel{\mathpalette\@versim<}}
\def\gsim{\mathrel{\mathpalette\@versim>}}
\def\@versim#1#2{\vcenter{\offinterlineskip
    \ialign{$\m@th#1\hfil##\hfil$\crcr#2\crcr\sim\crcr } }}
\def\boxit#1{\vbox{\hrule\hbox{\vrule\kern3pt
      \vbox{\kern3pt#1\kern3pt}\kern3pt\vrule}\hrule}}

\def\etal{{\it et. al.}}
\def\r#1{$\bf#1$}
\def\rb#1{$\bf\overline{#1}$}
\def\ty{{\widetilde Y}}
\def\t1{{\tilde 1}}
\def\ov{\overline}
\def\F{\widetilde F}
\def\Fb{\widetilde{\bar F}}
\def\Fbp#1#2{\widetilde{\bar F^#1_{#2}}}
\def\JL{J. L. Lopez}
\def\DVN{D. V. Nanopoulos}

\def\eV{\,{\rm eV}}

\def\GeV{\,{\rm GeV}}
\def\TeV{\,{\rm TeV}}
\def\y{\,{\rm y}}

\def\NPB#1#2#3{Nucl. Phys. B {\bf#1} (19#2) #3}
\def\PLB#1#2#3{Phys. Lett. B {\bf#1} (19#2) #3}

\def\PRD#1#2#3{Phys. Rev. D {\bf#1} (19#2) #3}
\def\PRL#1#2#3{Phys. Rev. Lett. {\bf#1} (19#2) #3}
\def\PRT#1#2#3{Phys. Rep. {\bf#1} (19#2) #3}
\def\MODA#1#2#3{Mod. Phys. Lett. A {\bf#1} (19#2) #3}
\def\IJMP#1#2#3{Int. J. Mod. Phys. A {\bf#1} (19#2) #3}
\def\TAMU#1{Texas A \& M University preprint CTP-TAMU-#1}

\nref\RB{K. Inoue, \etal, Prog. Theor. Phys. 68 (1982) 927; J. Ellis, \DVN, and
K. Tamvakis, \PLB{121}{83}{123}; J. Ellis, J. Hagelin, \DVN, and K. Tamvakis,
\PLB{125}{83}{275}; L. Alvarez-Gaum\'e, J. Polchinski, and M. Wise,
\NPB{221}{83}{495}; L. Iba\~n\'ez and C. L\'opez, \PLB{126}{83}{54} and
\NPB{233}{84}{545}; C. Kounnas, A. Lahanas, \DVN, and M. Quir\'os,
\PLB{132}{83}{95} and \NPB{236}{84}{438}.}
\nref\ttbar{V. Miransky, M. Tanabashi, and K. Yamawaki, \MODA{4}{89}{1043}
and \PLB{221}{89}{177}; W. Bardeen, C. Hill, and M. Lindner,
\PRD{41}{90}{1647}; W. Marciano, \PRL{62}{89}{2793}.}
\nref\ttbarsusy{W. Bardeen, T. Clark, and S. Love, \PLB{237}{90}{235};
M. Carena, \etal, Fermilab preprint FERMILAB-PUB-91/96-T (1991).}
\nref\TC{For a review see E. Farhi and L. Susskind, \PRT{74}{81}{277}.}
\nref\BT{S. Samuel, \NPB{347}{90}{625}; A. Kagan and S. Samuel, City College
preprints CCNY-HEP-90/17 and CCNY-HEP-91/8.}
\nref\TCx{D. Kennedy and P. Langacker, \PRL{65}{90}{2967} and
\PRD{44}{91}{1591}; A. Ali and G. Degrassi, DESY preprint DESY 91-035 (1991);
R. Casalbuoni, \etal, Univ. de Geneve preprint UGVA-DPT1991/07-735.}
\nref\KLNPY{J. Ellis and F. Zwirner, \NPB{338}{90}{317}; M. Drees and M.M.
Nojiri, KEK preprint KEK TH-290 (1991); K. Inoue, M. Kawasaki, M. Yamaguchi,
and T. Yanagida, Tohoku University preprint TU-373 (1991);
S. Kelley, \JL, \DVN, H. Pois, and K. Yuan, \TAMU{67/91} (To appear in Phys.
Lett. B).}
\nref\EKN{J. Ellis, S. Kelley and D. V.  Nanopoulos, \PLB{249}{90}{441},
\PLB{260}{91}{131}, and CERN preprint CERN-TH.6140/91;
U. Amaldi, W. de Boer, and H. F\"urstenau,
\PLB{260}{91}{447}; P. Langacker and M.-X. Luo, \PRD{44}{91}{817}.}
\nref\noscale{E. Cremmer, S. Ferrara, C. Kounnas, and \DVN, \PLB{133}{83}{61};
J. Ellis, A. B. Lahanas, \DVN, and K. Tamvakis, \PLB{134}{84}{429};
J. Ellis, C. Kounnas, and \DVN, \NPB{241}{84}{406} and \NPB{247}{84}{373}.
For a review see A. B. Lahanas and D. V. Nanopoulos, \PRT{145}{87}{1}.}
\nref\GSW{M. B. Green, J. H. Schwarz, and E. Witten, {\it Superstring Theory},
(Cambridge Univ. Press, 1987).}
\nref\JE{J. Ellis, Lectures given at Trieste Spring School and Workshop on
Superstrings, Trieste, Italy, Apr 11-22, 1988.}
\nref\Ginsparg{P. Ginsparg, \PLB{187}{87}{139}.}
\nref\Kaplu{V. Kaplunovsky, \NPB{307}{88}{145}.}
\nref\Lacaze{I. Antoniadis, J. Ellis, R. Lacaze, and \DVN, CERN preprint
CERN-TH.6136/91 (To appear in Phys. Lett. B).}
\nref\thresholds{S. Kalara, \JL, and \DVN, \TAMU{46/91} (To appear in Phys.
Lett. B).}
\nref\GO{For a review see \eg, P. Goddard and D. Olive, \IJMP{1}{86}{303}.}
\nref\ELNa{J. Ellis, J. Lopez, and \DVN, \PLB{245}{90}{375}.}
\nref\FIQ{A. Font, L. Ib\'a\~nez, and F. Quevedo, \NPB{345}{90}{389}.}
\nref\Lewellen{D. Lewellen, \NPB{337}{90}{61}; J. A. Schwarz,
\PRD{42}{90}{1777}.}
\nref\SMCY{B. Greene, K.H. Kirklin, P.J. Miron, and G.G. Ross,
\PLB{180}{86}{69};
\NPB{278}{86}{667}; \NPB{292}{87}{606}; S. Kalara and R.N. Mohapatra,
\PRD{36}{87}{3474}; J. Ellis, K. Enqvist, \DVN, and K. Olive,
\NPB{297}{88}{103}; R. Arnowitt and P. Nath, \PRD{39}{89}{2006}.}
\nref\SMOrb{L. Ib\'a\~nez, H. Nilles, and F. Quevedo, \PLB{187}{87}{25};
L. Ib\'a\~nez, J. Mas, H. Nilles, and F. Quevedo, \NPB{301}{88}{157};
A. Font, L. Ib\'a\~nez, F. Quevedo, and A. Sierra, \NPB{331}{90}{421}.}
\nref\SMFFF{A. Faraggi, \DVN, and K. Yuan, \NPB{335}{90}{347};
A. Faraggi, \TAMU{49/91}.}
\nref\revamp{I. Antoniadis, J. Ellis, J. Hagelin, and \DVN, \PLB{231}{89}{65}.}
\nref\PS{I. Antoniadis, G. Leontaris, and J. Rizos, \PLB{245}{90}{161};
T.T. Burwick, R.K. Kaiser, and H.F. Muller, \NPB{362}{91}{232};
D. Bailin, E.K. Katechou, and A. Love, Sussex preprint SUSX-TH-90-12 (1990).}
\nref\comp{D. S\'en\'echal, \PRD{39}{89}{3717}.}
\nref\Barr{S. Barr, \PLB{112}{82}{219}, \PRD{40}{89}{2457}; J. Derendinger,
J. Kim, and \DVN, \PLB{139}{84}{170}.}
\nref\AEHN{I. Antoniadis, J. Ellis, J. Hagelin, and \DVN, \PLB{194}{87}{231}.}
\nref\AEHNab{I. Antoniadis, J. Ellis, J. Hagelin, and \DVN, \PLB{205}{88}{459}
and \PLB{208}{88}{209}.}
\nref\KLN{S. Kalara, J. Lopez, and \DVN, \PLB{245}{90}{421},
\NPB{353}{91}{650}.}
\nref\decisive{J. L. Lopez and \DVN, \PLB{251}{90}{73}.}
\nref\RT{J. Rizos and K. Tamvakis, \PLB{251}{90}{369}.}
\nref\ELNpd{J. Ellis, \JL, and \DVN, \PLB{252}{90}{53}.}
\nref\STAB{J. L. Lopez and \DVN, \PLB{256}{91}{150}.}
\nref\sharp{\JL\ and \DVN, \TAMU{27/91} (To appear in Phys. Lett. B).}
\nref\conden{S. Kalara, \JL, and \DVN, \TAMU{69/91}.}
\nref\FFF{I. Antoniadis, C. Bachas, and C. Kounnas, Nucl. Phys. B
{\bf 289} (1987) 87; I. Antoniadis and C. Bachas, Nucl. Phys. B {\bf298} (1988)
586; H. Kawai, D.C. Lewellen, and S.H.-H. Tye, Phys. Rev. Lett. {\bf57} (1986)
1832; Phys. Rev. D {\bf34} (1986) 3794; Nucl. Phys. B {\bf288} (1987) 1;
R. Bluhm, L. Dolan, and P. Goddard, Nucl. Phys. B {\bf309} (1988) 330;
H. Dreiner, J. L. Lopez, D. V. Nanopoulos, and D. Reiss, Nucl. Phys. B
{\bf 320} (1989) 401.}
\nref\books{See \eg, {\it String theory in four dimensions}, ed. by M. Dine,
(North-Holland, 1988); {\it Superstring construction}, ed. by B. Schellekens
(North-Holland, 1989).}
\nref\FN{A. Faraggi and \DVN, \TAMU{78/90}.}
\nref\HFM{J. Lopez and \DVN, \NPB{338}{90}{73}.}
\nref\DSW{M. Dine, N. Seiberg, and E. Witten, \NPB{289}{87}{589}.}
\nref\Anoma{M. Dine, I. Ichinose, and N. Seiberg, \NPB{293}{87}{253};
J. Atick, L. Dixon, and A. Sen, \NPB{292}{87}{109};
M. Yamaguchi, H. Yamamoto, and T. Onogi, \NPB{327}{89}{704};
M. Yamaguchi, T. Onogi, and I. Ichinose, \IJMP{5}{90}{479};
M. Dine and C. Lee, \NPB{336}{90}{317}.}
\nref\LRT{G. Leontaris, J. Rizos, and K. Tamvakis, \PLB{243}{90}{220}.}
\nref\germans{J. Lauer, D. L\"ust, and S. Theisen, \NPB{304}{88}{236}.}
\nref\LT{G. Leontaris and K. Tamvakis, \PLB{260}{91}{333}.}
\nref\LNc{G. Leontaris, \PLB{207}{88}{447}; G. Leontaris and \DVN,
\PLB{212}{88}{327}; G. Leontaris and K. Tamvakis, \PLB{224}{89}{319};
S. Abel, \PLB{234}{90}{113}.}
\nref\LV{G. Leontaris and J. Vergados, \PLB{258}{91}{111}.}
\nref\neutrinomass{J. F. Wilkerson, Talk given at the 14th International
Conference on Neutrino Physics and Astrophysics ``Neutrino 90"
(CERN, Geneva, 1990), to appear.}
\nref\UY{S. Kelley, \JL, and \DVN, \TAMU{79/91}.}
\nref\MP{B. Greene, K. Kirklin, P. Miron, and G. Ross, \PLB{180}{86}{69},
\NPB{274}{86}{574}, B {\bf278} (1986) 667, B {\bf292} (1987) 606;
M. Bento, L. Hall, and G. Ross, \NPB{292}{87}{400}.}
\nref\ENR{J. Ellis, \DVN, and S. Rudaz, \NPB{202}{82}{69};
B. Campbell, J. Ellis, and \DVN, \PLB{141}{84}{229};
K. Enqvist, A. Masiero, and \DVN, \PLB{156}{85}{209}; R. Arnowitt,
A. Chamseddine, and P. Nath, \PRD{32}{85}{2348}.}
\nref\EHNT{J. Ellis, J. Hagelin, \DVN, and K. Tamvakis, \PLB{124}{83}{484}.}
\nref\Aspects{J. Ellis, J. Hagelin, S. Kelley, and \DVN, \NPB{311}{88/89}{1}.}
\nref\IMB{IMB-3 Collaboration, \PRD{42}{90}{2974}.}
\nref\cryptons{J. Ellis, J. Lopez, and \DVN, \PLB{247}{90}{257}.}
\nref\PDG{Particle Data Group, \PLB{239}{90}{1}.}
\nref\neutron{A. Gould, B. Draine, R. Romani, and S. Nussinov,
\PLB{238}{90}{337}.}
\nref\SY{A. Schellekens and S. Yankielowicz, \NPB{327}{89}{673};
\IJMP{5}{90}{2903}.}
\nref\Sch{A. Schellekens, \PLB{237}{90}{363}.}
\nref\sarkar{J. Ellis, G. Gelmini, \JL, \DVN, and S. Sarkar, CERN preprint
CERN-TH.5853/90.}
\nref\DKL{L. Dixon, V. Kaplunovsky, and J. Louis, \NPB{355}{91}{649}.}
\nref\FILQ{A. Font, L. Ib\'a\~nez, D. L\"ust, and F. Quevedo,
\PLB{245}{90}{401}.}
\nref\Jan{J. Louis, SLAC preprint SLAC-PUB-5527 (1991).}
\nref\ANT{I. Antoniadis, K. S. Narain, T. Taylor, \PLB{267}{91}{37}.}
\nref\Der{J. Derendinger, S. Ferrara, C. Kounnas, and F. Zwirner,
CERN preprint CERN-TH.6004/91;
G. Lopes Cardoso and B. Ovrut, University of Pennsylvania preprint
UPR-0464T (1991).}
\nref\Ig{I. Antoniadis, \PLB{246}{90}{377}.}
\nref\price{I. Antoniadis, J. Ellis, S. Kelley, and \DVN, CERN preprint
CERN-TH.6169/91.}
\nref\ILR{L. Ib\'a\~nez, D. L\"ust, and G. Ross, CERN preprint
CERN-TH.6241/91.}

\leftline{\titlefont TEXAS A\&M UNIVERSITY}
\leftline{\bf CENTER FOR THEORETICAL PHYSICS}
\Title{\vbox{\baselineskip12pt\hbox{CTP--TAMU--76/91}\hbox{ACT--49}}}
{Flipped SU(5): Origins and Recent
Developments{$^*$}\footnote{}{{$^*$}To appear in the
Proceedings of the 15th Johns Hopkins Workshop on Current Problems in
Particle Theory, Johns Hopkins University, August 26--28, 1991.}}
\centerline{JORGE~L.~LOPEZ\footnote{$^\dagger$}{Supported
by an ICSC--World Laboratory Scholarship.}
\footnote{$^\ddagger$}{Conference Speaker.} and D.~V.~NANOPOULOS}
\bigskip
\centerline{\CTPa}
\centerline{\CTPb}
\centerline{and}
\centerline{\HARCa}
\centerline{\HARCb}
\vskip .3in
\centerline{ABSTRACT}
We present an account of the early developments that led to the present
form of the flipped $SU(5)$ string model. We focus on the method used to
decide on this particular string model, as well as the basic steps followed
in constructing generic models in the free fermionic formulation of
superstrings in general and flipped $SU(5)$ in particular. We then describe
the basic calculable features of the model which are used to obtain its
low-energy spectrum: doublet and triplet Higgs mass matrices, fermion Yukawa
matrices, neutrino masses, and the top-quark mass. We also review the status
of proton decay in the model, as well as the hidden sector bound states called
cryptons. Finally, we comment on the subject of string threshold corrections
and string unification.
\bigskip
\Date{September, 1991}

\newsec{The Road to Superstring Models}
It is generally believed that the Standard Model (SM) of the strong and
electroweak interactions is to be viewed as an effective gauge theory valid
at energies below the electroweak symmetry breaking scale. Besides the usual
arguments in favor of more fundamental theories which encompass and
potentially explain the SM, there is a consistency requirement that must
be satisfied by any extension of the SM. This is related to the experimentally
observed breaking of the electroweak symmetry. At least two classes of
mechanisms come to mind to effect this breaking: the Higgs mechanism of
ordinary point-field theories induced by radiative corrections in the presence
of softly broken supersymmetry \RB, and dynamical symmetry breaking schemes
based on condensates of known (\eg, $t\bar t$
condensates \refs{\ttbar,\ttbarsusy}) or unknown
(\eg, technicolor \TC) fermions which mimic the elementary Higgs boson. One of
the clear advantages of the former mechanism is that the gauge hierarchy
problem is automatically solved, while supersymmetry may still be needed
in the latter since the condensation scales need to be rather
large.\foot{In fact, such a scenario has been explored recently in the
literature \refs{\BT,\ttbarsusy}.}

Experimentally speaking, there is mounting evidence against
certain class of composite Higgs theories \TCx. However, the main
drawback of these theories is more general. Due to their reliance on unknown
nonperturbative dynamics, these theories are not very well understood and of
limited calculability, and therefore their status as physical theories is
questionable. On the other hand, point-field supersymmetric theories are
perfectly calculable and their grand unified extensions highly predictive
\KLNPY. Furthermore, increasingly more precise measurements of the low-energy
gauge couplings and their extrapolation to very high energies shows a
remarkable
unification picture in the minimal unified model only when low-energy
superpartners are present \EKN.

Once supersymmetry is acknowledged as a major building block of modern unified
theories, the dynamical question of why the scale of supersymmetry breaking
is $\lsim1\TeV$ arises. This question is addressed naturally in no-scale
supergravity theories \noscale\ where a flat classical potential insures the
vanishing of the
cosmological constant (even after supersymmetry breaking). The last piece
of the puzzle is quantum gravity, and this problem has only one known solution,
namely superstring theory \GSW. With all this in mind we set out to construct a
unified supersymmetric superstring model. This task is however non-trivial
due to the immense classical vacuum degeneracy of string theory. We need
to make some educated choices.

\newsec{How to Pick a String Model}
It turns out that there is one property of string models which is surprisingly
restrictive: to gut or not to gut? \JE\ By which we mean, do we choose the
effective low-energy theory below the Planck mass to be a unified theory or
a at most abelian extension of the SM? Unified in this context means that at
least $SU(3)_C$ and $SU(2)_L$ are inside a bigger non-abelian gauge group.
This distinction is necessary since in string theory {\it all} gauge
couplings of non-unified gauge groups ``unify" \Ginsparg\ at the string
unification scale $M_{SU}$ \refs{\Kaplu,\Lacaze,\thresholds}, even though no
new degrees of freedom get excited at this scale (besides string massive
modes).

There are two types of symmetry breaking mechanisms in string models:
(a) at the string scale through the use of Wilson lines (\eg,
$E_8\times E_8\to SU(3)^3\times E_8$; $SO(44)\to SU(5)\times
U(1)\times\cdots$),
and (b) at lower energies via the usual Higgs mechanism. Wilson lines are
used in all string models to break the large primordial gauge symmetry down
to the initial high-energy gauge group of the model at scales $\approx M_{Pl}$.
{}From there on unified models with suitable Higgs sectors can break down to
the SM at intermediate scales. All these initial high-energy gauge groups are
reflections of two-dimensional symmetries called Kac-Moody algebras which
are parametrized by a positive integer called the ``level" \GO. Level-one
realizations have the property that no adjoint matter fields are present in
the spectrum \refs{\ELNa,\FIQ}. Higher-level realizations allow adjoint matter
representations as well as many more higher-dimensional representations
\refs{\ELNa,\FIQ}.
However, model-building using these higher-level algebras is technically a
rather difficult enterprise \Lewellen.

The choices are then clear:
\item{(1)} Construct Wilson-line breaking models with gauge group
${\rm SM}\times U(1)$'s at the string scale. These models can be built using
level-one or higher-level Kac-Moody algebras, although the latter are not
really needed. Several examples of this class have been constructed in the
literature \refs{\SMCY,\SMOrb,\SMFFF}.
\item{(2)} Construct models with unified gauge symmetry at the string scale
which need adjoint Higgs fields for symmetry breaking, using higher-level
Kac-Moody algebras. Realistic models of this type are beset with
constraints \refs{\ELNa,\FIQ}, although some examples exist \Lewellen.
\item{(3)} Construct models with unified gauge symmetry at the string scale
which do not need adjoint Higgs fields for symmetry breaking, using level-one
Kac-Moody algebras. An archetypal example of this class of models is flipped
$SU(5)$ \revamp, although other examples exist \refs{\PS,\comp}.
\medskip
Clearly the flipped $SU(5)$ model constructed under class (3) above is not
the unique string model. However, it certainly is the most developed string
model to date. It remains to be seen whether the proponents of any of the other
string models could eventually overcome some of the calculational or
phenomenological problems that their models may have, so that they too could
be brought to a level of development comparable to flipped $SU(5)$.

\newsec{Flipped SU(5): Introduction and Historical Remarks}
Group theoretically speaking, flipped $SU(5)$ is just an alternative
embedding of $SU(5)\times U(1)$ into $SO(10)$ and as such its basic
predictions for $\sin^2\theta_w$ and the proton decay lifetime have been
known for a while in its non-supersymmetric version \Barr. The main point
is that the electric charge generator $Q$ is only partially embedded in
$SU(5)$, \ie, $Q=T_3-{1\over5}Y'+{2\over5}\ty$, where $Y'$ is the $U(1)$
inside $SU(5)$ and $\ty$ is the one outside. This property affects the usual
unification picture as follows: the low-energy values of $\alpha_3$ and
$\alpha_2=\alpha/\sin^2\theta_w$ unify at a scale $M_X$, \ie, $\alpha_3(M_X)=
\alpha_2(M_X)=\alpha_5(M_X)\equiv\alpha_X$. The third coupling
$\alpha_1={5\over3}\alpha/\cos^2\theta_w$ gets related to $\alpha_X$ and
the $U(1)_\ty$ coupling at the scale $M_X$ by
\eqn\I{{25\over\alpha_1}={1\over\alpha_X}+{24\over\alpha_\ty}.}
Both $\alpha_\ty$ and $\alpha_5$ evolve further up to $M_{SU}$ where they
finally unify into $SO(10)$. This fact is used to fix the normalization
of the $\ty$ charge in much the same way that the embedding of $U(1)_{Y'}$
in $SU(5)$ is used to obtain the well-known factor of $5\over3$.

The flipped electric charge relation implies that the matter fields in each
generation are assigned to the $SU(5)$ representations as follows
\eqna\II
$$\eqalignno{\ov{\bf5}&=\left\{u^c,{{\nu_e\choose e}}\right\},&\II a\cr
                {\bf10}&=\left\{d^c,{{u\choose d}},\nu^c\right\},&\II b\cr
                {\bf1}&=\{e^c\},&\II c\cr}$$
clearly ``flipped" relative to the usual assignments. The neutral components
of the \r{10} and \rb{10} representations of Higgs fields are then used to
effect the unique symmetry breaking of $SU(5)\times U(1)$ down to
$SU(3)\times SU(2)\times U(1)$ without the use of adjoint Higgs
representations.

There are two very nice features of flipped $SU(5)$ model building \AEHN\ which
are seldom found in regular unified models: (i) a natural solution to the
doublet ($H$)-triplet ($D$) splitting problem of the Higgs pentaplets $h$
through the trilinear coupling of
Higgs fields: ${\bf10}_H\cdot{\bf10}_H\cdot{\bf5}_h\to\vev{\nu^c_H}d^c_HD$, and
(ii) an automatic see-saw mechanism to get heavy right-handed neutrino masses
through coupling to singlet fields $\phi$, ${\bf10}_f\cdot\ov{\bf10}_H\cdot\phi
\to\vev{\nu^c_{\ov H}}\nu^c\phi$. The left-handed neutrino fields have the
same Yukawa couplings as the up-type quarks.

The field theory blueprint that the string flipped $SU(5)$ model hoped to
emulate was proposed in 1987 \AEHN. It took two years and three papers
\AEHNab\ to produce the so-called revamped flipped $SU(5)$ model \revamp,
in the summer of 1989. This string model was indeed close to its field theory
analog but had the major advantage that its cubic superpotential could actually
be calculated (as opposed to just being postulated). This model was derived
in the free fermionic formulation of four-dimensional strings (see next
section)
and has the gauge group $SU(5)\times U(1)\times U(1)^4\times SO(10)_h\times
SU(4)_h$ at the string scale, where $SO(10)_h$ and $SU(4)_h$ are (semi)hidden
gauge groups. The superpotential of the model was promising, but there remained
several unanswered questions which were wishfully abscribed to uncalculated
higher-order terms in the superpotential. Among these were the full set of
doublet-triplet couplings, the determination of the number of light Higgs
doublets and the mixing between them, the hierarchy of quark and lepton masses,
and the elimination of unwanted fields from the light spectrum.

A major advance in free fermionic string model building came with the
elucidation of the techniques to calculate higher-order terms in the
superpotential \KLN. This breakthrough allowed a thorough investigation
of the structure of the model
\refs{\decisive,\RT,\ELNpd,\STAB,\sharp,\thresholds,\conden} and gave answers
to all the the above lingering questions. Two low-energy variants of the
model have since been constructed, differing in the assumed pattern of
$SU(5)\times U(1)$ symmetry breaking, as discussed below. We now describe
the calculational framework of the free fermionic formulation where the
revamped flipped $SU(5)$ model has been constructed.

\newsec{The Free Fermionic Formulation}
A reflection of the plethora of classical string vacua are the many
``formulations" one can use to construct these vacua (\ie, ``string models").
Basically all possible ways of getting string models are now known. The
two-dimensional (2d) world-sheet theory can only be conformal invariant at the
quantum level if extra 2d degrees of freedom are introduced besides the 4d
coordinates and their 2d superpartners. These extra ingredients can be
represented generally by appropriate conformal field theories with their own
set of 2d fields. The initial choice was to introduce additional spacetime
dimensions, as in the ten-dimensional Calabi-Yau models \GSW. This choice has
the advantage of yielding basically one 10d gauge group ($E_8\times E_8$).
However, the compactification process to 4d has never been fully understood.
The free fermionic formulation \FFF\ chooses extra free 2d fermions as the
supplementary degrees of freedom, thus no compactification is needed. The
drawback is that the number of models that can be constructed this way is
very large. Many other choices exist \books, but for calculational purposes the
free fermionic formulation is the most convenient one.

Free fermionic models are determined by the boundary conditions of the 2d
fermions as they loop around the 2d one-loop worldsheet (a torus). These
boundary conditions are arranged in ``vectors" of phases (\eg, periodic,
antiperiodic, etc.). Strict consistency requirements (2d supersymmetry,
2d modular invariance, etc) constrain the set of allowed vectors, and also the
possible states in the Hilbert space through generalized GSO projections.

The Hilbert space of physical states is completely calculable and the states
are built using standard 2d field theory tools (\ie, creation and
annihilation operators acting on a degenerate (Ramond sector) or non-degenerate
(Neveu-Schwarz sector) Fock vacuum). Each vector in the model generates a
sector
of states (which may or may not be projected out by the chosen GSO
projections).
It also generates a set of GSO projections on the states already present in
the model. Only after all desired vectors have been accounted for can one be
sure of the final spectrum of the model. The mass formula finally selects the
massless states. From this information all interactions in the model (\eg,
cubic and higher-order superpotential, D-terms, etc.) can be calculated \KLN.

Even though there
are no general rules, with some practice one can determine a minimal set of
vectors that produce a model with N=1 spacetime supersymmetry and has three
chiral families of quarks and leptons \FN. A particularly desired final
spectrum may or may not be achievable and is mostly the result of a tedious
trial-and-error process. Computer codes have been developed to this end \comp.
As an example of these vectors and their role, we show in Table I the ones
used in the revamped version of the flipped $SU(5)$ model. Their roles are
as follows: the vectors \r{1},$S$, and $\zeta$ give an N=4 spacetime
supersymmetric model with gauge group $SO(28)\times E_8$. Adding $b_1,b_2,b_3$
we achieve N=1 supersymmetry and $SO(28)\to SO(10)\times SO(6)^3$. Adding
$b_4,b_5,2\alpha$ breaks $SO(6)^3\to U(1)^6$ and $E_8\to SO(16)$. Finally
adding $\alpha$ breaks $SO(10)\to SU(5)\times U(1)$, $U(1)^6\to U(1)^4$, and
$SO(16)\to SO(10)\times SO(6)$.
\topinsert
\vbox{\tenpoint\noindent {\bf Table I}: The set of basis vectors used in the
construction
of the revamped flipped $SU(5)$ model. The various entries correspond to the
22 left-moving (left of the colon) and 44 right-moving 2d fermions. A 1 (0)
stands for periodic (antiperiodic) boundary conditions, and 1/2 for a phase
of $e^{i\pi/2}$. The symbol $A$ stands for
$A=({1\over2}{1\over2}{1\over2}{1\over2}1100)$ and $0_8=00000000$.}
\def\h{{\textstyle{1\over2}}}
\smallskip
\hrule
$$\eqalign{S&=(11\ 100\ 100\ 100\ 100\ 100\ 100\ :\ 000000\ 000000\
                                                        00000\ 000\ 0_8)\cr
b_1&=(11\ 100\ 100\ 010\ 010\ 010\ 010\ :\ 001111\ 000000\ 11111\ 100\ 0_8)\cr
b_2&=(11\ 010\ 010\ 100\ 100\ 001\ 001\ :\ 110000\ 000011\ 11111\ 010\ 0_8)\cr
b_3&=(11\ 001\ 001\ 001\ 001\ 100\ 100\ :\ 000000\ 111100\ 11111\ 001\ 0_8)\cr
b_4&=(11\ 100\ 100\ 010\ 001\ 001\ 010\ :\ 001001\ 000110\ 11111\ 100\ 0_8)\cr
b_5&=(11\ 001\ 010\ 100\ 100\ 001\ 010\ :\ 010001\ 100010\ 11111\ 010\ 0_8)\cr
\zeta&=(00\ 000\ 000\ 000\ 000\ 000\ 000\ :\ 000000\ 000000\ 00000\ 000\
                                                                1_8)\cr
\alpha&=(00\ 000\ 000\ 000\ 011\ 000\ 011\ :\ 000101\ 011101\ \h\h\h\h\h\
                                                        \h\h\h\ A)\cr}$$
\hrule
\endinsert
\newsec{The Flipped SU(5) String Model}
As explained in the previous section, to build a model in the free fermionic
formulation one needs to specify a consistent set of vectors and generalized
GSO projections. Well-defined rules then allow one to obtain the full massless
spectrum of the model which is shown in Table II.
Besides the usual gauge quantum numbers, string states possess internal 2d
degrees of freedom which appear in 4d as continuous or discrete global
symmetries. The latter are not shown in Table II but are obtained in the
process
and are essential in the calculation of the superpotential couplings \KLN. All
the cubic \refs{\revamp,\decisive,\RT}\ and quartic \decisive\ superpotential
couplings have been calculated and the nonvanishing quintic ones are tabulated.
For brevity here we just quote the cubic and quartic ones:
\eqna\III
$$\eqalignno{W_3=g\sqrt{2}\{&F_1F_1h_1+F_2F_2h_2+F_4F_4h_1
        +F_4\bar f_5\bar h_{45}+F_3\bar f_3\bar h_3+\bar f_1 l^c_1 h_1
                +\bar f_2 l^c_2 h_2\cr
        &+\bar f_5 l^c_5 h_2
                +{1\over\sqrt{2}}(F_4\bar F_5\phi_3+f_4\bar f_5\bar\phi_2
+\bar l^c_4 l^c_5\bar\phi_2)+\bar F_5\bar F_5\bar h_2+f_4\bar l^c_4\bar h_1\cr
        &+(h_1\bar h_2\Phi_{12}+h_2\bar h_3\Phi_{23}+h_3\bar h_1\Phi_{31}
                +h_3\bar h_{45}\bar\phi_{45}+{\rm h.c.})\cr
        &+{1\over2}(\phi_{45}\bar\phi_{45}+\phi^+\bar\phi^++\phi^-\bar\phi^-
                +\phi_i\bar\phi_i+h_{45}\bar h_{45})\Phi_3
                        +(\phi_1\bar\phi_2+\bar\phi_1\phi_2)\Phi_4\cr
        &+(\phi_3\bar\phi_4+\bar\phi_3\phi_4)\Phi_5
                +(\Phi_{12}\Phi_{23}\Phi_{31}+\Phi_{12}\phi^+\phi^-
                +\Phi_{12}\phi_i\phi_i+{\rm h.c.})\cr
&+D^2_1\bar\Phi_{23}+D^2_2\Phi_{31}
        +D^2_4\bar\Phi_{23}+D^2_5\bar\Phi_{31}
        +{1\over\sqrt{2}}D_4 D_5\bar\phi_3\cr
        &+T^2_1\bar\Phi_{23}+T^2_2\Phi_{31}+T^2_4\Phi_{23}+T^2_5\Phi_{31}
        +{1\over\sqrt{2}}T_4T_5\phi_2\cr
        &+{1\over\sqrt{2}}(\F_1\Fb_2\phi_4+\F_2\Fb_1\phi_1)+\F_2\Fb_2\phi^+
        +{1\over2}\F_3\Fb_3\Phi_3\cr
        &+\F_5\Fb_5\bar\Phi_{12}+\F_3\F_6 D_1+\F_3\Fb_4l^c_2\},
                                                        &\III a\cr}$$
$$\eqalignno{W_4=&c_1\,F_1\bar f_1\bar h_{45}\phi_1
        +c_2\,F_2\bar f_2\bar h_{45}\bar\phi_4
        +c_3(\bar f_2 f_4+l^c_2\bar l^c_4)\F_6\Fb_1
        +c_4(\bar f_5 f_4+l^c_5\bar l^c_4)\F_2\Fb_1\cr
        &+c_5(\bar f_3 f_4+l^c_3\bar l^c_4)T_3T_4
        +c_6\,F_3\bar F_5 D_3 D_5
        +(c_7\,\phi_2\bar\phi^-+c_8\,\bar\phi_2\phi^+)\F_4\Fb_4\cr
        &+(c_9\,\phi_3\bar\phi^-+c_{10}\,\bar\phi_3\phi^+)\F_6\Fb_6
        +c_{11}\,D_1D_5\F_1\Fb_4+c_{12}\,D_4D_5\F_1\Fb_2,&\III b\cr}$$
where the coefficients $c_i$ are given by
$$\eqalignno{c_1&=-3.07g^2,\quad c_2=ic_1,\quad c_3=-1.032ig^2,
                \quad c_4=2c_3,\cr
c_5&=0.487ig^2,\quad c_6=-0.259g^2,\quad c_7=-0.211ig^2,\quad
c_8=-0.829ig^2,\cr
c_9&=0.418ig^2,\quad c_{10}=0.245ig^2,\quad c_{11}=0.216g^2,
                        \quad c_{12}=g^2/2.     &\III c\cr}$$
Note that all couplings depend only on the unified string coupling $g$.
\topinsert
\baselineskip=12pt
{\tenpoint\noindent {\bf Table II}: The massless spectrum of the revamped
flipped $SU(5)$ model. The transformation properties of the observable sector
fields under $SU(5)\times U(1)$ are as follows: $F\,({\bf10},1/2)$,
$\bar f\,(\bar{\bf5},-3/2)$, $l^c\,({\bf1},5/2)$, $h\,({\bf5},-1)$. The hidden
sector fields transform under $SO(10)_h\times SU(4)_h$ as follows:
$T\,({\bf10},1)$, $D\,(1,{\bf6})$, $\F\,(1,{\bf4})$. The $\F_i,\Fb_j$ fields
carry $\pm1/2$ electric charges.}
\medskip
\hrule\smallskip
\noindent Observable Sector:
$$\eqalign{F_1&\quad \bar f_1\quad l^c_1\qquad h_1\quad \bar h_1\cr
        F_2&\quad \bar f_2\quad l^c_2\qquad h_2\quad \bar h_2\cr
        F_3&\quad \bar f_3\quad l^c_3\qquad h_3\quad \bar h_3\cr
        F_4&\quad       f_4\quad l_4\qquad h_{45}\quad \bar h_{45}\cr
        \bar F_5&\quad \bar f_5\quad l^c_5\cr}$$
\noindent Singlets:
$$\eqalign{&\phi_{45}\quad \bar\phi_{45}\qquad \Phi_{12}\quad \bar\Phi_{12}
                                        \qquad \Phi_{1,2,3,4,5}\cr
        &\phi^+\quad\bar\phi^+\qquad \Phi_{23}\quad\bar\Phi_{23}\cr
        &\phi^-\quad\bar\phi^-\qquad \Phi_{31}\quad\bar\Phi_{31}\cr
        &\phi_{1,2,3,4}\quad\bar\phi_{1,2,3,4}\cr}$$
\noindent Hidden Sector:
$$\eqalign{T_1\qquad D_1\qquad &\F_1\quad \Fb_1\cr
T_2\qquad D_2\qquad &\F_2\quad \Fb_2\cr
T_3\qquad D_3\qquad &\F_3\quad \Fb_3\cr
T_4\qquad D_4\qquad &\F_4\quad \Fb_4\cr
T_5\qquad D_5\qquad &\F_5\quad \Fb_5\cr
                &\F_6\quad \Fb_6\cr}$$
\hrule
\endinsert
A characteristic of this class of models is the presence of an anomalous
$U_A(1)$ in the gauge group, which is a linear combination of all the $U(1)$'s
in the model with nonvanishing trace. It can be shown \HFM\ that if there are
$n$ such $U(1)$'s, then the anomalous combination is given by
$U_A=k\sum_i[{\rm Tr}\,U_i]U_i$, where the coefficient $k$ is generally not
important but for properly normalized $U(1)$'s it is given by
$1/k^2=\sum_i[{\rm Tr}\,U_i]^2$. The remaining $n-1$ orthogonal linear
combinations are traceless and otherwise arbitrary. This phenomenon arises
because we are only considering the massless sector of a theory with an
infinite number of massive states. An effective low-energy theory of the
massless modes is not really anomalous since there is a built-in mechanism
\DSW\ that cancels all the potential anomalies in triangle graphs. However,
this state of affairs has observable consequences. Indeed, the D-term of the
anomalous $U_A(1)$ receives a one-loop contribution proportional to
${\rm Tr}\, U_A$ \Anoma, and the ``anomalous" gauge boson receives a two-loop
order mass. Since we want to preserve unbroken supergravity at $M_{Pl}$, the
D-terms of all $U(1)$'s must vanish, as well as all the F-terms, that is
\eqna\VI
$$\eqalignno{\vev{W}&=\vev{{\partial W\over\partial\phi_i}}=0,&\VI a\cr
        \vev{D_A}&=\sum_i q^i_A|\vev{\phi_i}|^2+\epsilon=0,&\VI b\cr
        \vev{D_a}&=\sum_i q^i_a|\vev{\phi_i}|^2=0,\quad a=1,2,3\,,&\VI b\cr}$$
where $\epsilon=g^2{\rm Tr}\,U_A/192\pi^2$. To restore the units one recalls
that $\kappa=\sqrt{8\pi}/M_{Pl}$ has been set to 1 in these equations. In the
revamped model ${\rm Tr}\,U_A=180$ and thus
$\epsilon=(7.4\times g\times10^{17}\GeV)^2$. Defining
$M\equiv M_{Pl}/2\sqrt{8\pi}\approx1.2\times10^{18}\GeV$, the ``natural" scale
of $\vev{\phi}$ is then $\vev{\phi}/M\sim1/10$. There are more unknowns than
equations to be solved, so the flatness condition cannot determine the vevs,
although some results actually follow \refs{\HFM,\RT}, such as
$\vev{\Phi_3,\Phi_{12},\bar\Phi_{12}}=0$. It can be shown \HFM\ that at least
three (the anomalous one and two anomaly-free ones) of the four $U(1)$'s are
broken by any given solution to the flatness conditions, and most known
solutions in fact break all four $U(1)$'s. The gauge group is then reduced to
$SU(5)\times U(1)\times SO(10)_h\times SU(4)_h$ at scales
$\sim\vev{\phi}\sim10^{17}\GeV$.

There are three built-in mass scales in the model: (i) the singlet vevs
$\vev{\phi}\sim10^{17}\GeV$, (ii) the $SU(5)\times U(1)$ breaking vevs
$V,\ov V\sim10^{16}\GeV$ (put in by hand), and (iii) the scale of hidden
sector condensation $\Lambda_{10}\sim10^{15}\GeV$ \LRT\ and
$\Lambda_4\sim10^{10-12}\GeV$ \LRT, determined dynamically. The ensuing
hidden matter condensates $\vev{TT},\vev{DD,\F\Fb}$ then provide another
source of effective masses. With all these mass scales and the all-orders
superpotential one can study the Higgs doublet and triplet mass matrices
and the fermion Yukawa matrices to determine the low-energy content of the
model. Before doing this, let us pause to appreciate the role of the
higher-order terms in the superpotential. A generic cubic $SU(5)\times U(1)$
invariant $\phi_1\phi_2\phi_3$ can receive contributions from all orders
of the form \STAB
\eqn\VII{c\,g^{N-2}\phi_1\phi_2\phi_3\,\vev{\phi}^n(V\ov V)^m\vev{TT}^p
        (\sqrt{2\alpha'})^{N-3},}
where $N=n+2m+2p+3$ and $c$ is an ${\cal O}(1)$ calculable constant. Using
\refs{\Ginsparg,\germans}\
$\kappa={1\over2}g\sqrt{2\alpha'}=\sqrt{8\pi}/M_{Pl}$, that is
$g\sqrt{2\alpha'}=1/M$, we get
\eqn\VII{c\,g\phi_1\phi_2\phi_3\,\left( {\vev{\phi}\over M}\right)^n
\left({V\ov V\over M^2}\right)^m
\left( {\vev{TT}\over M^2}\right)^p.}
Clearly, higher-order terms get naturally suppressed by powers of
$\vev{\phi}\sim10^{-1}$, $V\ov V/M^2\sim10^{-4}$,
$\vev{TT/M^2}\lsim(\Lambda_{10}/M)^2\sim10^{-6}$.

\newsec{The low-energy spectrum}
We now obtain the low-energy spectrum of the model by explicit examination
of the Higgs doublet and triplet mass matrices, the fermion Yukawa matrices,
and the neutrino see-saw matrix. We also run the third generation Yukawa
coulings to low energies to obtain predictions for $m_t$ and $\tan\beta$.
\subsec{Preliminaries}
We must first make an educated guess as to the pattern of $SU(5)\times U(1)$
breaking. The \r{10} of Higgs is in general a linear combination of $F_1,
F_2,F_3,F_4$. However, an analysis of the Higgs triplet mass matrix
shows \decisive\ that unless the vev is in $F_1$ or $F_2$ and/or $F_3$, there
will not be enough light $d^c$ states. The choice $F_1$ and/or $F_3$ is
preferred for the fermion mass spectrum. The two low-energy scenarios mentioned
above arise when: (i) $\vev{F_3}=0$ \decisive\ or (ii) $\vev{F_3}\not=0$
\sharp.

It so happens that these two choices were originally made together
with a choice for the way in which the extra $f_4,l_4$ matter states got
heavy. There are superpotential terms of the form: $f_4\sum_i\alpha_i\bar f_i$
and similarly for $l_4$. In Model (i) one took $\alpha_{1,2,5}\gg\alpha_3$,
whereas in Model (ii) the opposite was assumed. Below we will present results
for Model (ii) only since these are more interesting. (Besides, Model (i) may
suffer from too rapid proton decay \LT.)

It has not become clear until a recent study of the hidden matter condensates
in the model \conden\ that the assumptions about $\vev{F_3}$ and $\alpha_i$
which determine Models (i) and (ii), are actually compatible with each other.
The point is that $\alpha_3\propto\vev{T_3T_4}$ is non-vanishing and large
only if $\vev{F_3}\not=0$, and in this case $\alpha_3\gg\alpha_{1,2,5}$,
as advocated in Model (ii). If $\vev{F_3}=0$, $\vev{T_3T_4}\sim0$ and
$\alpha_3\ll\alpha_{1,2,5}$ as needed for Model (i). (However, the latter
choice
may lead to a pathological vacuum structure \conden.)
\subsec{Higgs doublet masses}
The analysis is complicated because of the several sources of Higgs doublet
masses, as follows
\eqna\VIII
$$\eqalignno{h_i\bar h_j&\to H_i\ov H_j,&\VIII a\cr
        F\bar f_i\bar h_j&\to \vev{\nu^c_H}L_i\ov H_j,&\VIII b\cr
        f_4\bar f_j&\to \bar L_4 L_j,&\VIII c\cr
        \bar F_5 f_4 h_i&\to \vev{\nu^c_{\ov H}} \bar L_4 H_i.&\VIII d\cr}$$
Note the potential mixing between ``Higgs" doublets $H_i$ and ``lepton"
doublets $L_j$. As remarked above, contributions of these types can
arise at any order and one must be certain that the emerging structure of the
matrix is indeed stable up to sufficiently high orders so that yet
higher-order contributions are negligible. As it turns out, remarkable
results hold which make the structure indeed stable \STAB, with many entries
remaining uncorrected to all orders in nonrenormalizable terms. The reasons
behind these results are the internal symmetries of the string states mentioned
above. In passing, let us quote the following result \STAB
\eqn\IX{\phi^N\equiv0,\qquad N\ge4,}
where $\phi^N$ is an arbitrary product of $N$ singlets. This amazing result
is very important since it implies that the F-flatness conditions only get
contributions from the cubic $\phi^3$ couplings, that is, they are stable.

The analysis of the doublet matrix can be done in steps. First we consider
only the all-orders contribution generated by singlet vevs. These leave
$H_1$, $H_{245}=\cos\theta H_2-\sin\theta H_{45}$, and
$\ov H_{12}=\cos\bar\theta\,\ov H_1-\sin\bar\theta\,\ov H_2$, $\ov H_{45}$
light, where $\tan\theta=\vev{\Phi_{23}}/\vev{\phi_{45}}$ and
$\tan\bar\theta=\vev{\Phi_{31}}/\vev{\bar\Phi_{23}}$. It can be shown that
$\vev{TT}$-generated effective mass terms give more structure to the remaining
light Higgs doublet mass matrix, with $H_1$ and $\ov H_{45}$ remaining light
and a mixing term $\mu H_1\ov H_{45}$, with
$\mu=\vev{\bar\phi_{45}}\vev{TT}/M^2$ \refs{\decisive,\sharp}.
If $\vev{\bar\phi_{45}}=0$ at $M_{Pl}$ and it grows a vev $\sim10^{11}\GeV$
after supersymmetry breaking, then $\mu\sim10^3\GeV$.
\subsec{Higgs triplet masses}
Analogously one can study the doublet-triplet splitting matrix and obtain
the Higgs triplet masses \refs{\decisive,\STAB}. The calculation is cumbersome
but three important results follow: (i) the structure is stable as before,
(ii) one must be careful with how the \r{10} vev is distributed among the
decaplets, and (iii) there is a Higgs triplet with mass $\sim10^{10-11}\GeV$.
\subsec{Fermion Yukawa couplings}
After analyzing the doublet and triplet Higgs mass matrices we are ready to
make
the identification of the quarks and leptons with the specific string
representations. This is a unique choice in Model (ii) \sharp:
\eqna\X
$$\eqalignno{t\,b\,\tau\,\nu_\tau&:\quad Q_4\ d^c_4\ u^c_5\ L_1\ l^c_1,&\X a\cr
c\,s\,\mu\,\nu_\mu&:\quad Q_2\ d^c_2\ u^c_2\ L_2\ l^c_2,&\X b\cr
u\,d\,e\,\nu_e&:\quad Q_\beta\ d^c_\beta\ u^c_1\ L_5\ l^c_5,&\X c\cr}$$
where the $\beta$ subscript refers to
$F_\beta\propto-\vev{F_3}F_1+\vev{F_1}F_3$
which is the linear combination that does not get a vev. With this
identification the Yukawa couplings can be written down immediately \sharp:
\eqna\XI
$$\eqalignno{\lambda_t&=\lambda_b=\lambda_\tau=g\sqrt{2},&\XI a\cr
\lambda_c&=3.04g\vev{\bar\phi_4}/M,\quad
\lambda_s=\lambda_\mu=g\{\sum_{i=1}^4c_{si}\vev{\bar\phi_i}^2+c_{s\pm}
                        \vev{\bar\phi^+\bar\phi^-}\}/M^2,&\XI b\cr
\lambda_u&=0,\quad \lambda_d=(V_3/V)^2,\quad
\lambda_e=g\{c_{e1}\vev{\bar\phi_1}^2+c_{e4}\vev{\bar\phi_4}^2+c_{e\pm}
\vev{\bar\phi^+\bar\phi^-}\}/M^2,\cr
                                &&\XI c\cr}$$
Some quintic couplings had to be calculated explicitly to obtain
the result $\lambda_s=\lambda_\mu$.

These Yukawa couplings at $M_{Pl}$ still need to be evolved down to low
energies with a few uncertainties along the way, such as the effect of
supersymmetry breaking at high scales, the details of the $SU(5)\times U(1)$
breaking, the decoupling of the various heavy modes along the way, and the
usual low-energy questions about the values of the light quark masses.
Nevertheless, let us point out some interesting features of this set of
Yukawa couplings:
\item{(a)} The successful GUT relation $\lambda_b=\lambda_\tau$.
\item{(b)} The mass ratios $m_s/m_b\sim(\vev{\phi}/M)^2\sim1/30$ and
$m_d/m_b\sim(V_3/V)^2\sim1/600$ could plausibly be accomodated by
$\vev{\phi}/M\sim1/6$ and $V_3/V\sim1/25$.
\item{(c)} The successful (?) GUT relation $\lambda_s=\lambda_\mu$.
\item{(d)} The unexpected (successful (?)) relation $\lambda_c/\lambda_t
\sim(\lambda_e/\lambda_\tau)^{1/2}$ (if
$\vev{\bar\phi_1}\sim\vev{\bar\phi^+\bar
\phi^-}\sim0$).
\subsec{Neutrino masses}
The see-saw neutrino mass matrix in
flipped $SU(5)$ involves $\nu,\nu^c$, and singlet states
$\phi$ \refs{\AEHN,\LNc}, through the superpotential couplings:
$\lambda^{ij}_uF_i\bar f_j\bar h_{45}\ni\lambda^{ij}_u\nu^c_i\nu_j
\vev{\ov H^0_{45}}$;
$w_{ij}F_i\ov F_5\phi_j\ni w_{ij}\ov V \nu^c_i\phi_j$; and
$\mu_{ij}\phi_i\phi_j$. In the string model, the see-saw matrix turns out to be
$14\times14$
$\{\nu_{1,2,5},\nu^c_{\beta,2,4},\phi_{1,2,3,4},\bar\phi_{45},\Phi_{3,4,5}\}$.
An important point is that all the $\nu^c\nu^c$ entries vanish. However, there
are two-loop radiatively induced contributions of the form \LV\
$m_{\nu^c\nu^c}\approx(\alpha_X/4\pi)^2(3/16)\lambda_{d,s,b}M_X\approx
5\times10^8m_{d,s,b}$, which contribute decisively to the structure of the
see-saw matrix. If we choose $\vev{\bar\phi_2}=0$, then the matrix
breaks up into three blocks with only three light eigenvalues: $\nu_1,\,m_1=0$;
$\approx\nu_2,\,m_2=0$; and
$\approx\nu_5,\,m_5\approx m_t\sqrt{\lambda_d}\vev{TT}/M^2$. Hence (see
Eqs. \X{}) $\nu_\mu$ and $\nu_\tau$ are massless to this level of
approximation, and $\nu_e$ has a mass in the eV range, within the current
experimental limit $m_{\nu_e}<10\eV$ \neutrinomass.
\subsec{Top-quark mass predictions}
In principle one cannot determine any fermion mass since the ratio of vevs
$\tan\beta=v_2/v_1$ has not been determined dynamically. The correct
procedure would be to run all gauge and Yukawa couplings and demand adequate
electroweak symmetry breaking. Without doing this one can only give an upper
bound on $m_t$: $m_t=\lambda_t\sin\beta\times174<174\lambda_t\lsim174\GeV$
for $m_b=5.0\GeV$, since otherwise $\lambda_t$ would blow up before the
unification scale. As an encouraging sign, an explicit calculation like the
one outlined above for the $\lambda_t=\lambda_b=\lambda_\tau$ scenario
in minimal supersymmetric GUTs gives \UY\ $m_t\approx90-150\GeV$ and
$\tan\beta\approx25-45$ for $m_b=4.9\pm0.1\GeV$. Some of these points have
also been found to be compatible with adequate electroweak breaking.

\newsec{Proton Decay}
Baryon number violating operators of dimension four ($qqq/qql$) and five
($qqql$) are a generic menace to unified models. Basically $d=4$ operators
must be forbidden by some extra symmetry, such as matter parity \MP, and
acceptable Higgs-induced $d=5$ operators are allowed only in some regions of
the parameter space of supersymmetric unified models \ENR. Nonrenormalizable
interactions at the Planck scale may also produce $d=4,5$ baryon number
violating operators. An analysis in a generic $SU(5)$ supergravity model
\EHNT\ indicates that the nonrenormalizable point coupling $g_b$ must be
bounded by $g_b\lsim7\times10^{-25}\GeV^{-1}\ll1/M_{Pl}$. (In contrast,
in the minimal $SU(5)$ supersymmetric GUT model one finds
$g_b\sim G_Fm_cm_s\sin\theta_c/m_H$, which gives $M_H\gsim10^{17}\GeV$.)
Hence, if nonrenormalizable couplings are of order $g_b\sim 1/M_{Pl}$, as
one would naturally expect, then these new $d=5$ operators would be a disaster
giving $\tau_p\sim10^{20}\,{\rm y}$.

The various operators contributing to proton
decay in this model have to be studied carefully since some of them may be
potentially disastrous. These operators are:
\item{(a)}Nonrenormalizable terms in the superpotential of the
form \refs{\ELNpd,\LT}\ $FFF\bar f{1\over M}$ contain effective $d=5$ $qqql$
operators which could be very large \refs{\ELNpd,\LT}\ since the overall
coefficient
is expected \decisive\ to be ${\cal O}(1)$. These terms appear first at
fifth order \refs{\decisive,\LT}\ (multiplied by $\VEV{\phi}/M$) and always
contain the $\bar f_3$ field, and thus are harmless in Model (ii) since
$\bar f_3$ does not contain any light states.
\item{(b)}Dimension-five Higgsino mediated operators \Aspects\
are constructed via the usual tree-level diagrams involving three
superpotential couplings: $FFh$, $F\bar f\,\bar h$, and $h\bar h$. Since we
are interested in taking $\bar f$
to be $\bar f_{1,2,5}$, the associated $\bar h$ is always $\bar h_{45}$. The
needed mixing term $h\bar h_{45}$ is proportional to $\mu$ for $h_{1,2}$ and
to $\VEV{\bar\phi_{45}}$ for $h_3$, and the $F\bar f\,\bar h$ vertex coupling
is in effect the up-quark Yukawa matrix. A careful analysis shows
that the coefficients of the resulting effective $d=5$ operators are always
suppressed enough, although the potential proton decay rates could be close to
the current experimental limit.
\item{(c)}Dimension-six Higgs boson mediated operators could be important
because of the existence of ${\cal O}(10^{10-11})\GeV$ Higgs triplet states,
as mentioned above. There are two classes of diagrams \Aspects\ originating
from superpotential couplings of the forms:
$F\bar f\,\bar h/F^\dagger\bar f^\dagger\bar h^\dagger$ and $FFh/\bar f^\dagger
l^{c\dagger}h^\dagger$. The first class yield effective operators suppressed by
$1/\tan^2\beta$ relative to their counterpart in SUSY GUTs, while
the second class give negligible contributions.
\item{(d)}Dimension-six gauge boson exchange operators are sufficiently
suppressed in flipped $SU(5)$ models \ELNpd, where it is found that
$\Gamma(p\to\bar\nu\pi^+)=2\Gamma(p\to e^+\pi^0)=\Gamma(n\to e^+\pi^-)
=2\Gamma(n\to\bar\nu\pi^0)$, with
$\tau(p\to e^+\pi^0)
\approx3\times10^{31}\left({M_X\over10^{15}\GeV}\right)^4\y$. However, in the
present model the particle assignment in \X{}\ indicates that the dominant
decay modes are actually $p\to\bar\nu_\tau\pi^+$ and $n\to\bar\nu_\tau\pi^0$.
For comparison, the present experimental
lower limit is $\tau(p\to e^+\pi^0)>5.5\times10^{32}\y$ \IMB.

\newsec{Cryptons}
The hidden sector of the model contains the 22 matter fields given in Table II.
The electric charge generator $Q$ is given by $Q=T_3-{1\over5}Y'+{2\over5}\ty$,
where $U(1)_{Y'}$ is the $U(1)$ inside
$SU(5)\supset SU(3)_C\times SU(2)_L\times U(1)_{Y'}$. Since $Q$ is not
completely embedded in a simple group, the possibility exists for
electrically charged exotic states with no $SU(5)$ quantum numbers,
such as the $\F_i,\Fb_j$ above which have $Q=\pm1/2$ \refs{\ELNa,\cryptons}.
Since light free
fractionally charged particles are not observed \PDG, and their existence
could have grave astrophysical consequences \neutron, there must exist
a mechanism to either make them superheavy or somehow bind them into
neutral bound states (as in QCD).

The solution to the charged quantization problem can be encoded in the
following experimentally motivated charge quantization dogma: All (massless)
fractionally charged particles must have nontrivial quantum numbers under
unbroken nonabelian gauge groups, such that when confinement sets in and thus
only gauge singlets are observable, the resulting physical states are
integrally charged. In a string-derived model this condition can be
conveniently implemented in the language of simple currents \SY\ of the
Kac-Moody algebra underlying the gauge group \refs{\Sch,\ELNa}. Compatibility
with the string rules places severe restrictions on the gauge groups and
their Kac-Moody levels, which could enforce such a quantization condition.
It can be shown \ELNa\ that the following charge quantization rule can be
consistently imposed on the spectrum of the flipped $SU(5)$ string model:
$\alpha={1\over3}t_3+Q+{1\over2}t_4+{1\over2}c\in{\bf Z}$, where $t_3, t_4$,
and $c$ are the triality, quadrality, and conjugacy classes of the respective
$SU(3)_C, SU(4)_h$, and $SO(10)_h$ representations. One can readily verify
that this condition is satisfied by all massless states in the model. Also,
one can show \cryptons\ that this holds at all massive levels as well.

Models which do not have a hidden sector sticky enough to confine the
existing fractionaly charged states are likely to be in trouble. On the
other hand, potentially realistic models that confine fractional charges
will then generally contain integer-charge ``hidden hadrons" as their
solution to the charge quantization problem. From the light spectrum of
the model three kinds of $SU(4)_h\times SO(10)_h$ invariant bound states
follow:
\eqn\XII{{\rm cryptons}\quad\cases{
{\rm hidden\ mesons:}&$T_iT_j,\,D_iD_j,\,\F_i\Fb_j,\quad(0,\pm1)$;\cr
{\rm hidden\ baryons:}&$\F_i\F_jD_k,\,\Fb_i\Fb_jD_k,\quad(0,\pm1)$;\cr
{\rm
tetrons:}&$\F_i\F_j\F_k\F_l,\,\Fb_i\Fb_j\Fb_k\Fb_l,\quad(0,\pm1,\pm2)$;\cr}
}
where we have indicated in parenthesis the possible electromagnetic charges
for each class of crypton. The superpotential for these hidden fields can
be calculated, and it is seen that most of these fields become superheavy.
However, there
are four fields that remain as light as their respective confinement scales:
$T_3,D_3,\F_{3,5},\Fb_{3,5}$. Since a renormalization group analysis \LRT\
indicates that $\Lambda_4\approx10^{11-12}\GeV$ and
$\Lambda_{10}\approx10^{14-15}\GeV$, we expect the $SU(4)_h$ bound states to
be much lighter than $SO(10)_h$ bound states in general. The lightest $SU(4)_h$
bound states are expected to be the hidden mesons $\F_{3,5}\Fb_{3,5}$ and
tetrons $\F^4_{3,5}$ and $\Fbp{4}{3,5}$. Because of its larger crypto-charge,
one would expect bound states of the \r{6} field $D_3$ to be somewhat heavier.
We expect the lightest hidden $SU(4)_h$ meson to be analogous to the
$\pi^0:\,\pi^0_4\approx(\F_3\Fb_3-\F_5\Fb_5)/\sqrt{2}$,
$m^2_{\pi^0_4}\approx\Lambda_4\times m_{\F_{3,5},\Fb_{3,5}}$,
with charged $\pi^\pm_4\approx(\F_3\Fb_5,\Fb_3\F_5)$ states slightly heavier
because of electromagnetic mass splitting analogous to that in QCD. Also,
as in QCD we expect the $\eta^0_4\approx(\F_3\Fb_3+\F_5\Fb_5)/\sqrt{2}$
state to be significantly heavier because of a $U_A(1)$ anomaly. By analogy
with the nucleon and $\Delta$ states in QCD, the lightest tetrons are
expected to be the neutral $\F^2_3\F^2_5$ and $\Fbp{2}{3}\Fbp{2}{5}$ states,
with singly-charged $\F^3_3\F_5,\F_3\F^3_5,\Fbp{3}{3}\Fb_5$, and
$\Fb_3\Fbp{3}{5}$ states somewhat heavier, and doubly-charged
$\F^4_3,\F^4_5,\Fbp{4}{3}$, and $\Fbp{4}{5}$ states even heavier.

By considering the likely crypton decays one can show that the lightest
neutral tetrons have lifetimes in excess of $10^{16}$ years, and hence are
possible candidates for the dark matter of the Universe. There are also
cosmological \cryptons\ constraints on metastable
cryptons. A naive estimate of the crypton relic abundance
$\Omega_C\approx(\Lambda_4/100\TeV)^2\gg1$ indicates the need for entropy
releasing mechanisms. There are several mechanisms which
dilute the crypton relic density to levels $\Omega_C\ll1$. However,
$\Omega_C\approx0.1-1$ is not ruled out. Long-lived cryptons also satisfy
all known constraints on massive unstable neutral relic particles \sarkar.

\newsec{String Unification}
String theory predicts the scale at which all gauge couplings should unify
to be $M_K\approx7.3\times g\times10^{17}\GeV$ \Kaplu. However, this scale must
be corrected to include the so-called string threshold corrections. These
arise in much the same way as in regular gauge theories when thresholds of
massive particles are crossed. The novelties in string theory are that the
threshold is only approached from below and that the massive states are
infinite
in number. The one-loop renormalization group equation for the gauge couplings
can be generally written as follows \Kaplu
\eqn\XIII{{16\pi^2\over g_a^2(\mu)}
                ={16\pi^2\over g^2}+b_a\ln{M_K^2\over\mu^2}+\Delta_a,}
where $\Delta_a$ are the string threshold corrections and $b_a$ is the
coefficient of the one-loop beta function. In generic orbifold (and free
fermionic) models these have been calculated to be \refs{\DKL,\FILQ}
\eqn\XIV{\Delta_a=-\sum_\alpha{{\textstyle{1\over2}}} b^{\alpha}_a
\ln\left[|\eta(iT_\alpha)|^4{\rm Re\,}T_\alpha\right]+c_a+Y,}
where $\eta$ is the Dedekind function and $b^\alpha_a$ are the beta function
coefficients of the three N=2 supersymmetry sectors into which the massless
spectrum can be split. Equivalently \Jan\ one can express $\Delta_a$ in terms
of the charges of the states under the modular transformations of the moduli
fields $T_\alpha$. These fields parametrize the size of the compatification
manifold. The constant $c_a$ is model-dependent, but in a large class of
models $c_a=b_a\cdot c$. This contribution is small in all known cases
\refs{\Kaplu,\Lacaze,\thresholds}\ and will be neglected in what follows. The
quantity $Y$ is irrelevant in practical applications but it is of some
theoretical interest \refs{\Jan,\ANT,\Der}.

A nice simplification occurs when one takes the values of the three moduli
to be the same, since then ${1\over2}\sum_\alpha b^\alpha_a=b_a$ \Ig\ implies
that \thresholds
\eqn\XV{\Delta_a\approx b_a\left[{\pi\over3}{\rm Re}\,T-\ln{\rm Re}\,T+
                                        {\rm small\ corrections}\right].}
In the case of flipped $SU(5)$, the free fermionic formulation has $T=1$ and
the string unification scale becomes
\eqn\XVi{M_U\approx M_Ke^{\pi/6}\approx1.24\times g\times10^{18}\GeV.}
The significance of this result cannot be over-emphasized. At this scale the
string couplings unify and start their running to low energies. Furthermore,
the scale is determined by the string unified coupling, which becomes the
only unknown in the theory. This unification scale is two orders of magnitude
larger than the one expected from a straightforward evolution of the low-energy
couplings in the minimal supersymmetric $SU(5)$ model \EKN. Put it differently,
if one starts at $M_U$ with a value of $g$ such that $\alpha_{em}$ at low
energies comes out right, then at low energies one obtains
$\sin^2\theta_w=0.218$ and $\alpha_3=0.20$, which are in gross disagreement
with their very precise experimental counterparts.

Two alternatives have been suggested to reconcile these results. One can
throw in extra vector-like matter representations at suitable intermediate
scales to delay unification \price, or one can exploit the $T_\alpha$
dependence of $\Delta_a$ to push $M_U$ down \ILR. A combination of both
possibilities is in principle possible, although the latter cannot happen
in free fermionic models \ILR. In the case of flipped $SU(5)$, things are
more complicated due to the non-minimal matter content at only approximately
known intermediate mass scales. However, it appears that the first alternative
will need to be pursued to obtain a satisfactory model. This modification of
the model is currently under investigation; we believe that the main features
of the model outlined above will remain for the most part intact.
\newsec{Conclusions}
We have described the early developments leading up to the flipped $SU(5)$
string model, focusing on the rationale behind this particular string model
and basic model-bulding in the free fermionic formulation of superstrings,
which was used to construct the flipped $SU(5)$ model. A study of the
all-orders
superpotential allows us to obtain the low-energy spectrum of the model, which
entails an analysis of the doublet and triplet Higgs mass matrices, the fermion
Yukawa matrices, and the see-saw neutrino matrix. We also showed that in spite
of several operators contributing to the proton decay rate, it is not a problem
in the latest version of the model. We also reviewed one of the possible
``smoking guns" of string, namely the hidden sector bound states called
cryptons. Finally, we commented on the subject of string threshold corrections
and string unification.
\bigskip
\bigskip
\noindent{\bf Acknowledgments}: This work has been supported in
part by DOE grant DE-FG05-91-ER-40633.
\listrefs
\bye